\documentclass[a4paper]{jpconf}
\usepackage{graphicx}
\usepackage[pdf]{pstricks}
\usepackage{epstopdf}

\begin{document}
\title{Mini-AOD: A New Analysis Data Format for CMS}

\author{G Petrucciani$^{1}$, A Rizzi$^{2}$ and C Vuosalo$^{3}$, \newline
on behalf of the CMS Collaboration}

\address{$^{1}$ CERN}
\address{$^{2}$ University of Pisa and INFN}
\address{$^{3}$ University of Wisconsin-Madison}

\begin{abstract}
The CMS experiment has developed a new analysis object format ("Mini-AOD")
targeting approximately 10\% of the size of the Run 1 AOD format. The
motivation for the Mini-AOD format is to have a small and quickly derived data
format from which the majority of CMS analysis users can start their analysis
work. This format is targeted at having sufficient information to serve about
80\% of CMS analysis, while dramatically simplifying the disk and I/O resources
needed for analysis. Such large reductions were achieved using a number of
techniques, including defining light-weight physics-object candidate
representations, increasing transverse momentum thresholds for storing
physics-object candidates, and reduced numerical precision when it is not
required at the analysis level. In this contribution we discuss the critical
components of the Mini-AOD format, our experience with its deployment and the
planned physics analysis flow for Run 2 based on the Mini-AOD.
\end{abstract}

\section{Introduction}
The Large Hadron Collider (LHC) at CERN continues to advance the exploration
of the energy frontier as it begins its new run, called Run 2, in 2015 by
colliding protons at a center-of-mass energy of 13 tera-electronvolts (TeV),
an energy never before achieved in the laboratory. \mbox{Run 1}, from 2010-2012, provided
the data that led to the discovery of the Higgs boson. It was followed by a two-year
period of upgrade work to prepare the collider and its associated detector experiments
for the new higher energy and the much larger amount of collision data.
During \mbox{Run 2},
the CMS experiment [1] is expected to collect ten times more data than it did
in \mbox{Run 1}. However, the disk storage resources for CMS will be increasing only
marginally. For \mbox{Run 1}, CMS physics analysis was based upon the Analysis
Object Data (AOD) format, but this format is too large for it to be used in
the same way for Run 2. At the end of 2013, the total size of AOD stored by CMS
for Run 1 was about 20 petabytes. Increasing this amount by ten times for Run 2
would vastly exceed available storage resources. 
To tackle this challenge, CMS has developed a new,
condensed data format called Mini-AOD.

\section{Mini-AOD and the CMS Data Flow}
\begin{figure}[tb]
\begin{center}
\includegraphics[width=35pc]{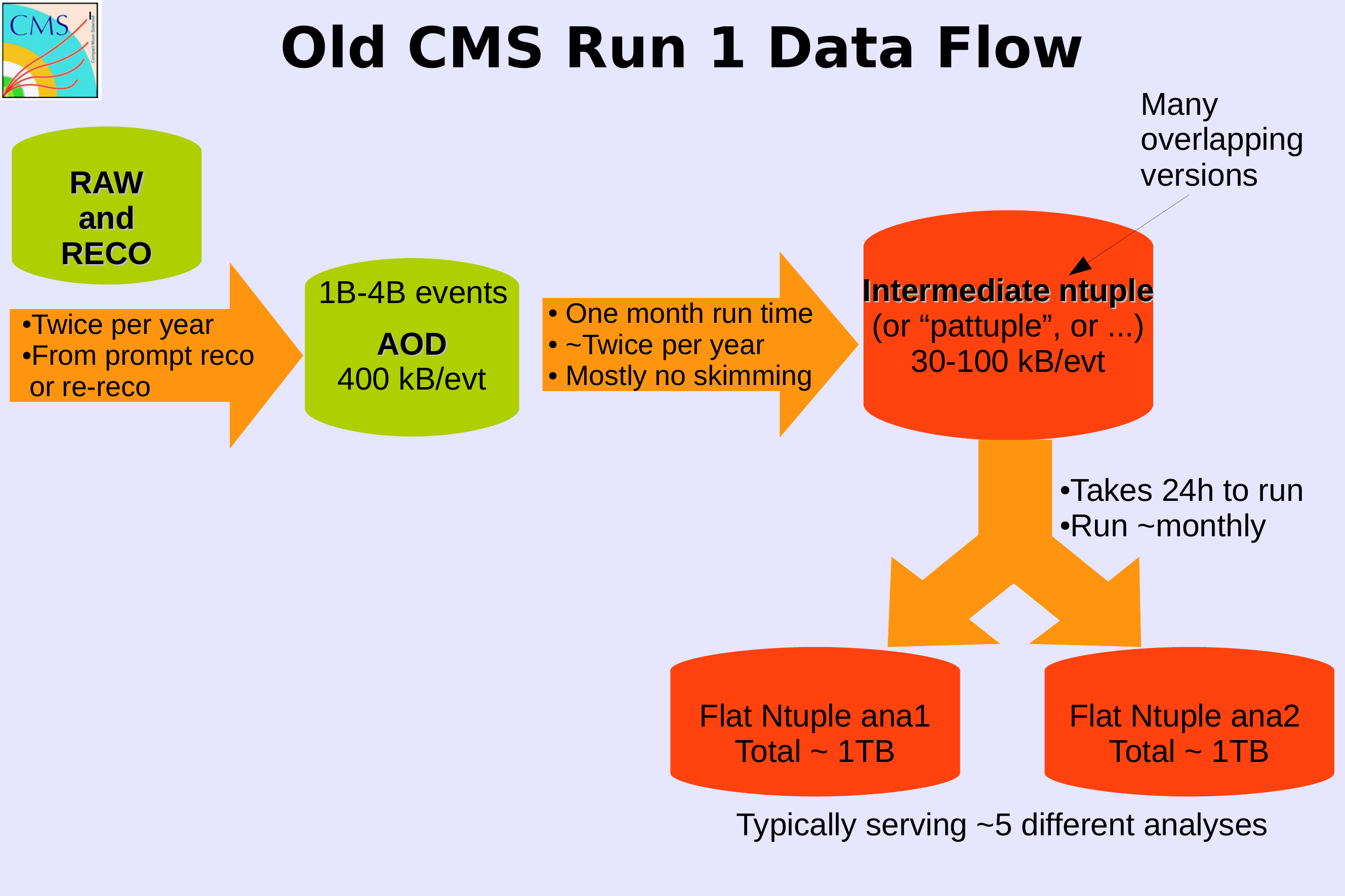}\hspace{2pc}
\end{center}
\caption{\label{dmrun1}Data flow for CMS Run 1. AOD comprises about one to
four billion events, with each event taking about 400 kilobytes.}
\end{figure}

For Run 1, the CMS data flow included several steps (Fig. \ref{dmrun1}). Raw data were digitized,
and collision events were reconstructed to form the Reco dataset. About twice
a year, the AOD dataset was generated from Reco. AOD comprised about one to
four billion events of roughly 400 kilobytes each in size. From multiple copies
of AOD spread across many CMS computing sites, 
physics analysis groups generated intermediate datasets called ntuples that
had more specialized formats intended for particular analysis needs.
The last step was for individual analysis groups to generate
simple, flat ntuples for their specific needs from the intermediate ntuples.
Unfortunately, the different sets of
intermediate ntuples were largely overlapping, and many
copies of them were made throughout the CMS storage infrastructure. In Run 2,
given the vast increase in the amount of data, this data flow would not be
sustainable.

\begin{figure}[htb]
\begin{center}
\includegraphics[width=35pc]{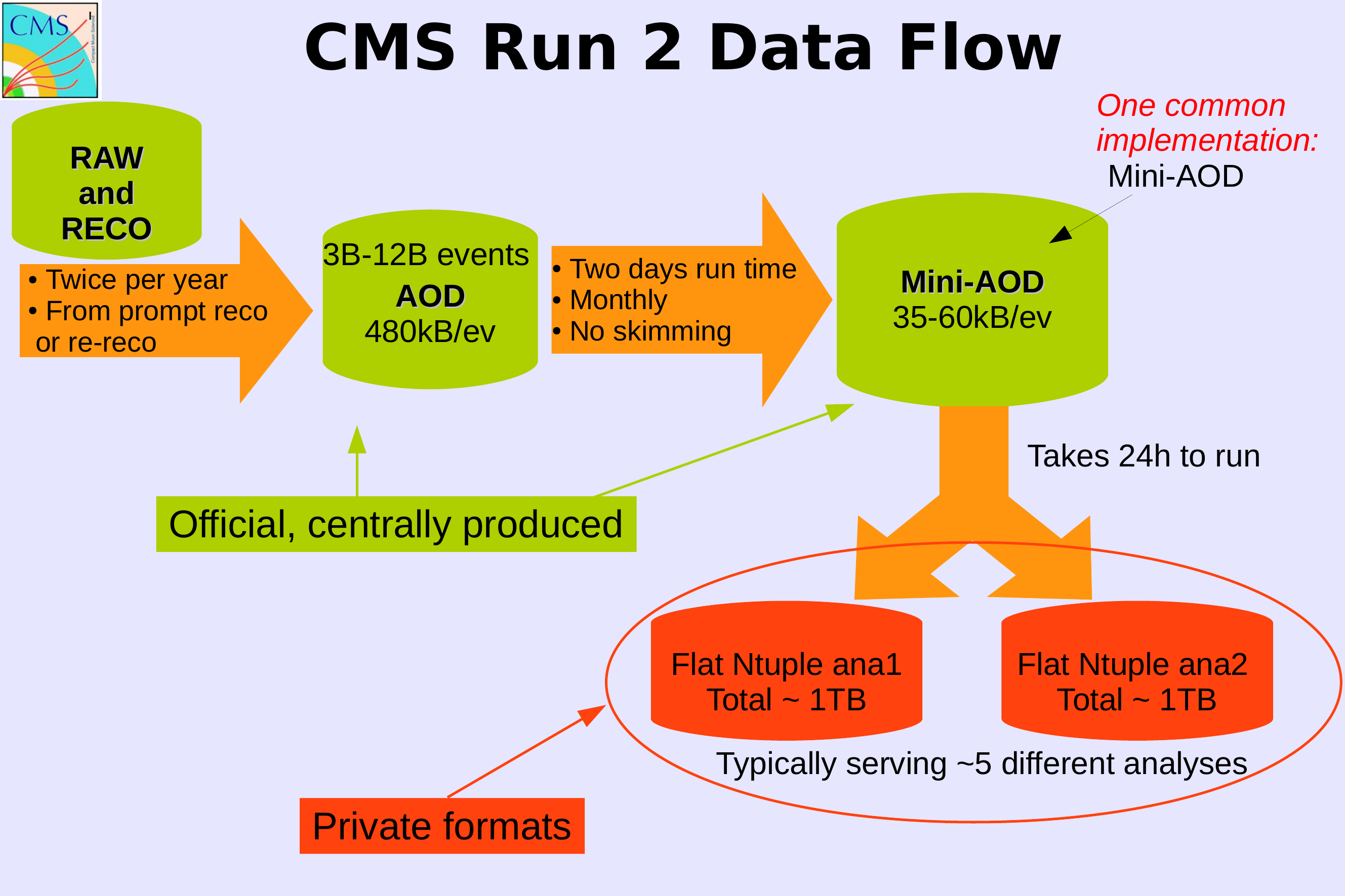}\hspace{2pc}
\end{center}
\caption{\label{dmrun2}Data flow for CMS Run 2. The size of AOD is about
480 kilobytes per event, while Mini-AOD is around 40-50 kilobytes per event.
Total AOD size will grow substantially compared to Run 1, going as high as
12 billion events or even more, depending on how much data is ultimately 
collected.}
\end{figure}

The new CMS data flow for Run 2 replaces the many, duplicative intermediate
ntuples with one standard, condensed dataset called Mini-AOD
(Fig. \ref{dmrun2}).
Without the need for specialized intermediate ntuples, 
the multiple copies of AOD used in Run 1 to support ntuple generation
have also been eliminated.
Mini-AOD is produced centrally by the CMS computing
group and provides a common foundation for CMS physics analyses.
Its compressed format is one tenth the size of AOD,
and it provides CMS the solution for
handling the large influx of data coming from Run 2. It is designed to meet
the needs of most CMS physics analyses while fitting within storage constraints.

\section{Mini-AOD Design Philosophy}
Mini-AOD was developed under the guidance of a basic design
philosophy with the following principles:

\begin{itemize}
\item Use the minimum amount of space.
\item Extract only the minimum required data from existing data formats.
\item Re-use existing data formats and algorithms where possible without
including unnecessary data.
\pagebreak
\item Maintain flexibility for:
\begin{itemize}
\item Supporting a wide variety of CMS physics analyses.
\item Novel analysis techniques.
\item Re-tuning of algorithms and calibrations that may be necessary after the
first new data at 13 TeV are analyzed.
\end{itemize}
\item Don't over-optimize. The basic requirements are to be able to store about
five billion events at a CMS computing center and to be able to process them in one
or two days. Once those requirements are met, further optimization and compression
of Mini-AOD is not necessary.
\end{itemize}

Efficient use of space is not the only objective Mini-AOD must meet. It must also
support the full range of CMS physics analyses, from standard model measurements to
searches for exotic new physics. And it must provide the flexibility to respond to
the challenges of collisions at the new, higher energy of 13 TeV. 
In particular, the number of additional collision interactions, called
pile-up, is much higher in Run 2 than in Run 1 and presents difficulties for accurately
reconstructing collision events efficiently.

\section{Content of Mini-AOD}
Based upon the design philosophy, the following components are included in
Mini-AOD. First it contains high-level physics objects: leptons, photons, jets,
and $E_{\mathrm{T}}^{\mathrm{miss}}$, which is defined as follows.
$E_{\mathrm{T}}^{\mathrm{miss}} =  |-\sum \vec{p}_{\mathrm{T}(i)}|$,
where $\vec{p}_{\mathrm{T}(i)}$ is $\vec{p}$ of 
reconstructed particle \textit{i} projected on the
plane perpendicular to the LHC beams.

\begin{figure}[htb]
\begin{center}
\includegraphics[width=35pc, trim=0 13pc 0 0,clip=true]{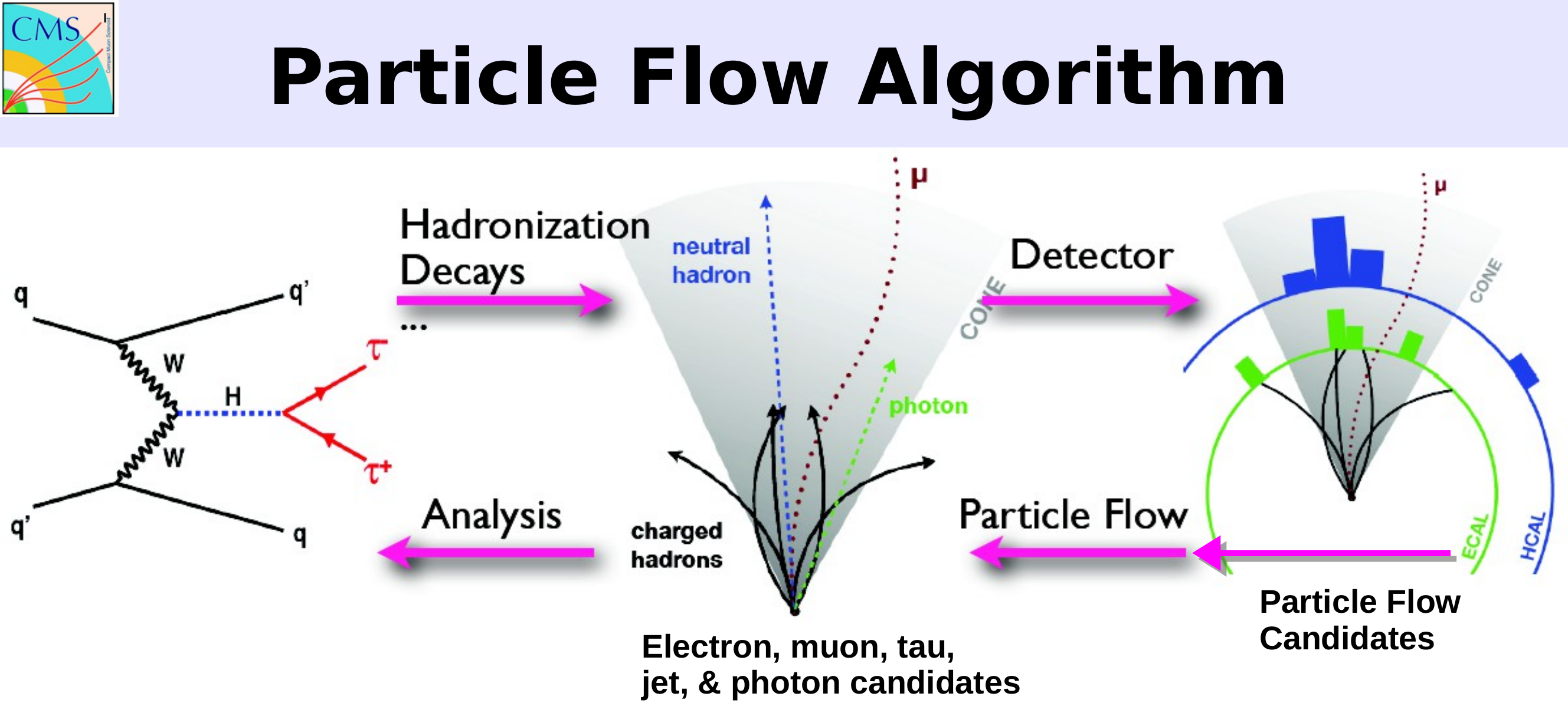}\hspace{2pc}
\end{center}
\caption{\label{pflow}CMS Particle Flow (PF) algorithm [2, 3]. The top half of the
diagram shows how collisions lead to particle decays and final state particles,
which leave tracks and deposits in the CMS detector. The bottom half shows that
PF candidates are derived from detector information and then become input for
the PF algorithm that uses them to construct high-level physics objects like
leptons and jets, which then are used by analysts to reconstruct the collision
event.}
\end{figure}

These objects are stored in the CMS Physics Analysis Tools format that includes detailed
information. Second, Mini-AOD contains all particle
candidates in a packed format that contains only basic kinematic information.
The candidates are produced by the CMS Particle Flow (PF)
reconstruction algorithm [2, 3] (Fig. \ref{pflow}), and their presence allows analysts to
re-reconstruct physics objects with new techniques directly from Mini-AOD.
Third, trigger information is also stored. Triggers are used to decide which
collisions events are recorded. The trigger data identify which triggers
were tripped and enables calculation of trigger efficiencies. Fourth, information
about simulated particles is included. Final state particles are stored, as
well as generated jets and reference information. Fifth, some miscellaneous
information like the interaction vertices and $E_{\mathrm{T}}^{\mathrm{miss}}$
filters is also included.

\section{Physics Objects in Mini-AOD}
Mini-AOD contains leptons, photons, jets, and $E_{\mathrm{T}}^{\mathrm{miss}}$ objects.
High-quality electrons are saved with detailed information if their transverse
momentum ($\mathrm{p}_{\mathrm{T}}$) is greater than 5 giga-electronvolts (GeV), or, if
not, with only their detector clusters. Full muon information is saved for muon candidates
that meet basic quality requirements. Likewise, high-quality tau candidates
with $\mathrm{p}_{\mathrm{T}} > 18$ GeV are saved. Similar to electrons, photon
candidates that pass quality requirements are stored in detail, while lower
quality photons are stored with only basic detector cluster information.

Two collections of jets are saved: one for general use,
and one intended for substructure studies. The general-use
jets must have $\mathrm{p}_{\mathrm{T}} > 10$ GeV and include b-quark jet discriminators
and secondary decay vertex information. Jets in the second collection
must have $\mathrm{p}_{\mathrm{T}} > 100$ GeV and include jet substructure
information. 

\section{Particle Flow Candidates in Mini-AOD}
PF candidates are the building blocks used by PF to reconstruct high-level
physics objects. Basic information for all PF candidates is included in
Mini-AOD: four-momentum, charge, impact parameters, and particle type (electron,
photon, etc.). Quality
flags relating to association to the interaction vertex and tracking hits in
the detector are also provided. Lossy compression is applied to the variables,
limiting precision to about 0.1\%. Compression is facilitated by sorting the
PF candidates, converting \texttt{double} variables (typically 8 bytes)  
to \texttt{float} (typically 4 bytes), and reducing precision of matrices
from $10^{-7}$ to $10^{-4}$.

The presence of all PF candidates in Mini-AOD supports many important physics
analysis tasks. Varied algorithms for pile-up mitigation can be studied, and
lepton and photon isolation algorithms can be
re-computed with these different pile-up algorithms. Reconstructed jets can
be re-clustered in substructure studies, and b-jet discrimination can be 
re-computed with new criteria. The PF candidates allow analysts to use Mini-AOD
to test and utilize a wide variety of candidate-based algorithms.

The physics objects and PF candidates are fully cross referenced, so each object
is linked to the PF candidates from which it was reconstructed. This linking
facilitates event interpretation and more accurate calculation of isolation.

\section{Simulated Particles}
For simulated events, only a selected set of particles is stored because the 
simulated particle format, called GenParticle, takes a lot of space. First,
a set called pruned GenParticles that includes initial partons, heavy flavor particles, electroweak
bosons, and leptons is stored in full. Second, a set called packed GenParticles
that have only the four-momentum and particle type is saved for particles representing
the final state particles in the event. Generated jets and some reference information
is also saved.

With the packed GenParticles, an analyst can re-make the generated jets with
various algorithms. The pruned GenParticles enable event classification, flavor
definition, and matching to reconstructed physics objects. Links from each
packed GenParticle to its last surviving ancestor
pruned GenParticle allow the decay chain of the event to be reconstructed.

\begin{figure}[htb]
\begin{center}
\includegraphics[width=35pc]{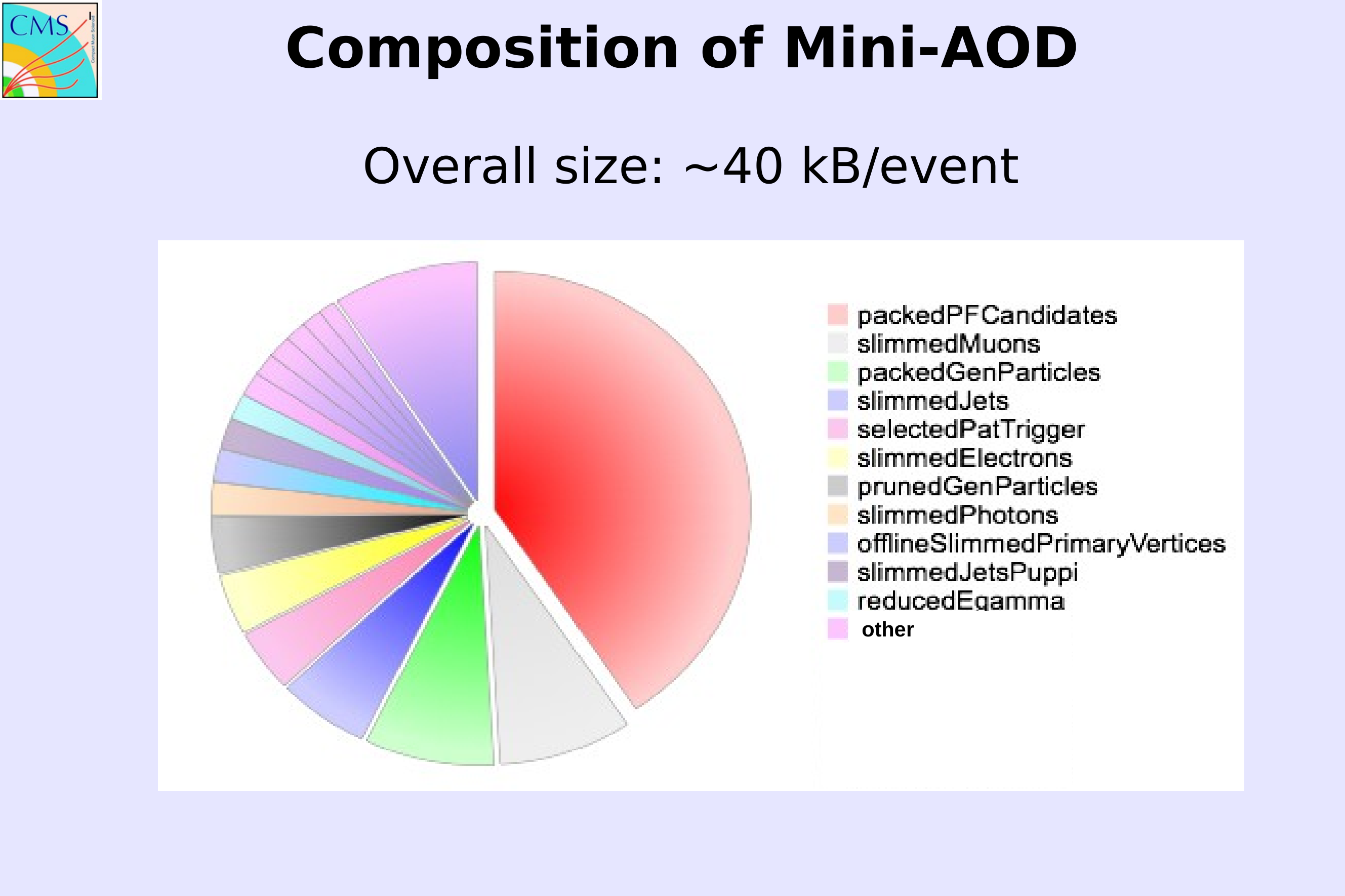}\hspace{2pc}
\end{center}
\caption{\label{macomp}Composition of Mini-AOD. The average size of an event
contained in Mini-AOD for simulated events involving top and antitop quark decay
is about 40 kilobytes.}
\end{figure}

Figure \ref{macomp} shows a breakdown of the composition of Mini-AOD.

\section{Development History}
Mini-AOD was conceived in 2014. February brainstorming sessions led to a
prototype in March. Further development and a series of integration test
releases occurred in April and May, and Mini-AOD was ready for production use
in June. The latter half of the year comprised two concentrated campaigns of
testing and validation, the Computing, Software and Analysis Challenge 2014 and
the Physics Challenge 2014. During these campaigns, many bugs and issues were
discovered and fixed. Now, as Run 2 begins in 2015, Mini-AOD has been validated
and is ready for the flood of new data.

\section{Summary}
The CMS experiment faces the challenge of analyzing ten times more data in the
new LHC \mbox{Run 2} starting this year compared to Run 1 in 2010-2012. However, CMS
data storage resources have increased only marginally. The Run 1 method of using
AOD as the foundation of physics analysis and storing many overlapping versions
of intermediate datasets is not sustainable for Run 2. For these reasons, a new
compressed data format called Mini-AOD has been developed. Mini-AOD is 10\% of
the size of AOD, and it replaces the multiple intermediate datasets used in Run 1.
It provides a standard foundation for CMS physics analysis. Its optimized collections
store only the minimum required amount of information, but it maintains
flexibility for a variety of new and customized analysis techniques and also
for any re-tuning of the analysis process that might be necessary as CMS performs
the ground-breaking exploration of the energy frontier at \mbox{13 TeV}. Mini-AOD has undergone
a lengthy and thorough testing and validation process and is now ready to meet
the challenge of LHC Run 2.

\section*{References}
\numrefs{99}
\item Chatrchyan S {\it et al} (CMS Collaboration) 2008 The {CMS} experiment at
the {CERN} {LHC}{\it JINST} vol 3 p S08004
\item {CMS Collaboration} 2009 Particle--flow event reconstruction in {CMS} and
performance for jets, taus, and {MET} {\it CMS Physics Analysis Summary}
CMS-PAS-PFT-09-001 (http://cdsweb.cern.ch/record/1194487)
\item {CMS Collaboration} 2010 Commissioning of the particle-flow event
reconstruction with the first {LHC} collisions recorded in the {CMS} detector
{\it CMS Physics Analysis Summary}
CMS-PAS-PFT-10-001 (http://cdsweb.cern.ch/record/1247373)

\endnumrefs

\end{document}